\def\year{2021}\relax
\documentclass[letterpaper]{article} %
\usepackage{aaai21}  %
\usepackage{times}  %
\usepackage{helvet} %
\usepackage{courier}  %
\usepackage[hyphens]{url}  %
\usepackage{graphicx} %
\usepackage{natbib}  %
\usepackage{caption} %
\title{Engaging Politically Diverse Audiences on Social Media}
\author {
    Martin Saveski, 
    Doug Beeferman,
    David McClure,
    Deb Roy \\
}
\affiliations{
    Massachusetts Institute of Technology \\
    \{msaveski, dougb5, dclure, dkroy\}@mit.edu
}

\usepackage{makecell}

\usepackage{color} 
\usepackage[dvipsnames]{xcolor}

\usepackage{amssymb}
\newcommand{\mybullet}{
    \noindent  
    $\triangleright$
}

\newcommand{\nl}{n_{\textit{{\scriptsize left}}}}
\newcommand{\nr}{n_{\textit{{\scriptsize right}}}}

\newcommand{\pl}{p_{\textit{{\scriptsize left}}}}
\newcommand{\pr}{p_{\textit{{\scriptsize right}}}}

\newcommand{\DeltaT}{\Delta_{\textit{{\scriptsize left}}}^{\textit{{\scriptsize T}}}}
\newcommand{\DeltaC}{\Delta_{\textit{{\scriptsize left}}}^{\textit{{\scriptsize C}}}}

\newcommand{\plT}{p_{\textit{{\scriptsize left}}}^{\textit{{\scriptsize T}}}}
\newcommand{\prT}{p_{\textit{{\scriptsize right}}}^{\textit{{\scriptsize T}}}}

\newcommand{\plC}{p_{\textit{{\scriptsize left}}}^{\textit{{\scriptsize C}}}}
\newcommand{\prC}{p_{\textit{{\scriptsize right}}}^{\textit{{\scriptsize C}}}}

\newcommand{\pwl}{p(\textit{word}~|~\textit{left-leaning})}
\newcommand{\pwr}{p(\textit{word}~|~\textit{right-leaning})}

\begin{document}

\maketitle

\begin{abstract}
We study how political polarization is reflected in the social media posts used by media outlets to promote their content online. In particular, we track the Twitter posts of several media outlets over the course of more than three years (566K tweets), and the engagement with these tweets from other users (104M retweets), modeling the relationship between the tweet text and the political diversity of the audience.  We build a tool that integrates our model and helps journalists craft tweets that are engaging to a politically diverse audience, guided by the model predictions. To test the real-world impact of the tool, we partner with the PBS documentary series Frontline and run a series of advertising experiments on Twitter. We find that in seven out of the ten experiments, the tweets selected by our model were indeed engaging to a more politically diverse audience, illustrating the effectiveness of our approach. 
\end{abstract}

\section{Introduction}
The U.S. news media is more politically fragmented than ever.  Americans of different political identities inhabit divergent media worlds, with ever more separation between the sources of information that they engage with and trust~\cite{pew2020election, starbird2017examining, garimella2018political}.  Media outlets have  an opportunity to counteract this polarization by actively  promoting their content in a way that brings in a more politically diverse audience. In this paper, we study how news content is promoted through the lens of political polarization, and we offer a model, editing tool, and evaluation framework which could potentially mitigate it.

Since about half of Americans get their news from social media~\cite{pew2020news}, we focus on how media outlets promote their content online.   Much work has been done on predicting the popularity of social media posts and identifying the content characteristics associated with popularity, including brevity~\cite{gligoric2019causal}, readability~\cite{tan2014effect}, emotional valence~\cite{berger2012makes}, and topic variation~\cite{aldous2019view}. Other studies have focused on the popularity of news, examining what makes news headlines more engaging. They find that headlines that are questions~\cite{lai2014makes}, contain signal words~\cite{kuiken2017effective}, sentimental words~\cite{rieis2015breaking}, and second-person pronouns~\cite{lamprinidis2018predicting} tend to be more engaging. Media outlets~are also increasingly testing multiple versions of their headlines, seeking to maximize the reach of their content~\cite{hagar2019optimizing, matias2019upworthy}. 
However, less attention has been paid to how the language that media outlets use to promote their content influences the engagement of different audiences, e.g., left- vs.\ right-leaning audiences. 

A recent line of work has focused on methods for automatically neutralizing subjective~\cite{pryzant2020automatically} or polarizing~\cite{chen2018learning,liu2021political} text. Generally speaking, these approaches work by first identifying subjective or polarizing text segments (words or sentences) and then substituting them with text segments that are semantically similar but less subjective or polarizing. While these techniques show very promising results, in practice, journalists need to follow specific editorial standards and require more flexibility in composing and editing content. Motivated by this observation, we develop tools that supplement the journalists’ writing process by providing instant feedback based on model predictions, and contextualizing the model outputs by highlighting keywords and surfacing similar historical content.  

Another challenge in studying how users respond to different social media posts is external validity. Researchers often have to rely on surveys administered in a different context (e.g., a crowd-sourcing platform vs.\ the social media platform itself) and reaching a potentially different set of participants. We attempt to overcome this challenge by running advertising experiments on the platform. While advertising experiments are not a panacea (Section~\ref{sec:ads-exps}), they allow us to test how thousands of users respond to different social media posts on the platform.

\textbf{The present work.}
In this paper, we 
(1) model the relationship between the content of promotional social media posts and the political diversity of the users who engage with them, (2) build tools that help journalists write posts that are engaging to more politically diverse audiences, and (3) test the effectiveness of our tools using advertising experiments on Twitter.\footnote{The code and data needed to replicate our analyses are available at: \url{https://github.com/msaveski/engaging-diverse-audiences}}

To investigate the relationship between the post content and audience diversity, we tracked all tweets posted by five major media outlets across the political spectrum (New York Times, CNN, Wall Street Journal, Fox News, Breitbart; sorted left to right) and Frontline over more than three years, collecting over 566K tweets and 104M retweets (Section~\ref{sec:data}). To measure the political diversity of each tweet’s audience, we compute the entropy of the distribution of left- vs.\ right-leaning retweeters. Then, we use this data to train machine learning models that, given the tweet text, predict the political diversity of the audience (Section~\ref{sec:predictive-modeling}).

To test our models in a real-world setting, we partner with the PBS documentary series Frontline. Frontline is a world-renowned investigative journalism program that produces in-depth documentaries on various domestic and international issues. Like other programs, Frontline uses social media to promote their films. However, as a PBS program, their goal is not just to maximize engagement but also to reach a politically diverse audience. We build a web application that integrates our models and helps Frontline’s journalists craft tweets that are engaging to a more politically diverse audience, guided by the model predictions (Section~\ref{sec:news-bridge-web-app}). 

To test the effectiveness of our tools, we run a series of advertising experiments promoting Frontline’s tweets (Section~\ref{sec:ads-exps}). In each experiment, we select a pair of tweets about the same documentary---one predicted to be engaging to a less and another predicted to be engaging to a more politically diverse audience---and measure the engagement of left- and right-leaning users with each tweet. We find that in seven out of the ten advertising experiments, the treatment tweets were indeed engaging to a more politically diverse audience, illustrating the potential of our tools.

\section{Data}
\label{sec:data}

\textbf{Data Collection.}
To systematically study the relationship between posts' content and audience diversity, we collect a large number of tweets posted by media outlets and data related to the users who engaged with those tweets. We use the text of media outlets' tweets to characterize the content and the user data to characterize the audience's political leaning. 

We tracked the tweets of five major media outlets and Frontline over more than three years, from January 2017 to March 2020. We selected the New York Times, CNN (left), Wall Street Journal (center), Fox News, and Breitbart (right) as they have large followings on Twitter, their tweets consistently receive a lot of engagement, and together they cover the full political spectrum~\cite{bakshy2015exposure,budak2016fair}. Due to the limit of the number of tweets that we could ingest per month, we were unable to consider a larger set of outlets. 

We collected 566K tweets and the corresponding 104M retweets. We note a drop in the volume of tweets posted by Fox News after Nov 8, 2019, when they stopped tweeting in protest against Twitter after a group of demonstrators posted the home address of Tucker Carlson, one of the network’s show hosts. To avoid any bias due to data censoring, we excluded from the analysis tweets posted over the last week of the data collection period, but counted the new retweets of tweets posted prior to that. 

\textbf{Measuring Tweet Audience Diversity.}
To measure the political diversity of the audience of each news tweet, we compute the entropy of the retweeters' political alignment scores.  We consider only retweets as they are a clear sign of agreement and endorsement of the content. We exclude so-called quote tweets, retweets with commentary, which can be used to express disagreement with the original tweet. To ensure that we have a reasonable estimate of the tweets’ audience characteristics, we filter out tweets with fewer than three retweeters whose political alignment score we could estimate. Next, we describe how we estimate the political alignment scores, and in Section~\ref{sec:predictive-modeling}, we discuss our decision to discretize these scores.

\textbf{Measuring User Alignment.}
To measure the users' political alignments, we analyze the links that they share in their tweets, retweets, and quote tweets. We build on previous work by Bakshy et al.~\shortcite{bakshy2015exposure}, which demonstrates that left- and right-leaning users share significantly different content. Based on their analysis---grounded in the users' self-reported political leaning---they released political alignment scores for the 500 most shared domains with scores ranging from -2 (left-leaning) to +2 (right-leaning).

To calculate the political alignment of each user, we extract the URLs of the content that they posted, look up the political alignment of each URL domain, and take the average. We classify users with negative average alignment scores as left-leaning and users with positive average alignment scores as right-leaning.  
To collect tweets posted by the users, we use a 3-year snapshot of the Twitter Decahose, which includes a 10\% sample of all public tweets posted between January 2017 and December 2019. 
We use a snapshot of the Decahose instead of the REST API to be able to retrieve more than 3,200 tweets allowed by the REST API and to have access to original instead of shortened tweet URLs. 
We consider only users who have shared at least three URLs.

\section{Political Alignment Evaluation}
\label{sec:political-alignment-score-evaluation}
Since our analyses extensively rely on the quality of the estimated political alignments, in this section, we thoroughly evaluate them by (1) comparing the composition of the users in our sample with the share of two-party votes in recent elections, (2) comparing with alignments estimated based on network information, and (3) running a survey on Mechanical Turk. 

\begin{figure}
\centering
\includegraphics[width=\linewidth]{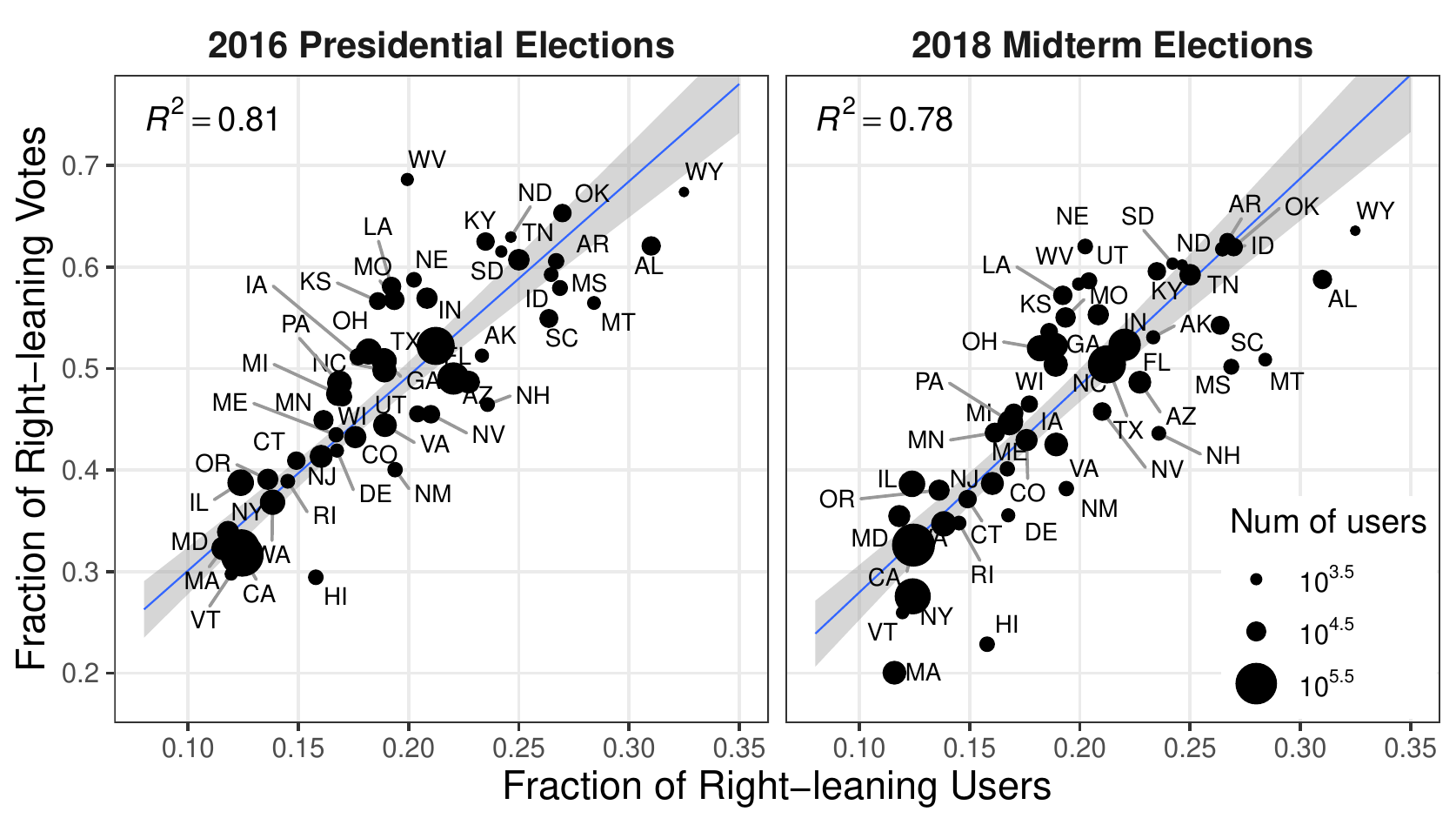}
\caption{Comparison between the proportion of right-leaning users in our dataset for each U.S. state and the proportion of right-leaning votes in the 2016 and 2018 elections.}
\label{fig:alignments-per-state}
\end{figure}

\textbf{User Alignments: Comparison with the Share of Two-Party Vote per State.}
To validate the user political alignments, we compare the proportion of users classified as right-leaning in each U.S. state against the Republican share of the two-party vote in the 2016 presidential and 2018 midterm elections, following a similar methodology as \citet{barbera2015tweeting} and \citet{demszky2019analyzing}.
We use a sample of 2.3M users from our Decahose snapshot who had shared at least three URL domains with known alignment and whose location we could infer based on the location field in their user profiles. As shown in Figure~\ref{fig:alignments-per-state}, we find a strong correlation between the fraction of users classified as right-leaning in our dataset and both the proportion of votes for Donald Trump in 2016 ($R^2 = 0.81$, in a weighted linear regression adjusting for the number of users per state) and the proportion of votes for Republican candidates for the House of representatives in 2018 ($R^2=0.78$).

\textbf{Tweet Alignments: Comparison with Network-Based User Alignments.}
Next, we compare the alignment scores calculated based on the users’ content sharing patterns with alignments inferred from the users’ following patterns~\cite{barbera2015tweeting}. The key idea behind this method is that users are more likely to follow accounts that align with their political views, including accounts with unambiguous leaning such as legislators.

For each tweet posted by the media outlets, we compute the average alignment of the retweeters based on their content sharing patterns and their following patterns. We find a very strong correlation between the average alignment scores of the retweeters computed using the two methods (Pearson $R~=~0.99$, $p<2.2e^{-16}$). However, in 93\% of the tweets, the alignments based on the content sharing patterns give us more information to estimate the tweets' alignment scores in our dataset. We can infer the user alignment scores of more retweeters per tweet and have enough information to compute the alignment scores of more tweets.

\textbf{Tweet Audience Diversity: Survey.}
To further validate our measure of audience diversity, we run a Mechanical Turk survey asking left- and right-leaning participants whether they would consider sharing sample tweets. We build on recent work demonstrating that self-reported willingness to share political news in online surveys conducted on Mechanical Turk correlates with actual sharing on Twitter~\cite{mosleh2020self}.

We sample ten tweets for each outlet and Frontline, excluding Fox News whose account was inactive at the time due to their boycott of Twitter. To ensure that the tweets in the survey were on current topics, we consider only tweets posted by the outlets during the six weeks before the survey. To ensure that the sample tweets capture the variation of tweet alignments, we take a stratified sample for each outlet: we first compute the deciles of the outlet's tweet alignment score distribution and then sample one tweet from each decile. We further ensure that the tweets do not require any additional context to understand and do not reveal the organization that posted them.

The survey includes 25 questions. We first explain to the participants that they will be shown a series of social media posts posted by major media outlets and that they will be asked whether they will consider sharing them. Then, we show them one tweet at a time and ask ``Would you consider sharing the following post on social media?'' We do not show the images or the headlines associated with the tweets. To avoid any bias due to ordering effects, we randomize the order of the questions and the order of the response buttons. To ensure high-quality responses, we invite only participants from the U.S. that have completed at least 100 tasks and have a high approval rate ($>$98\%). We administer the survey such that, for each sample tweet, we obtain 50 responses by self-identified ``Liberal'' and 50 responses by self-identified ``Conservative'' participants. We compensate the participants 60 cents or roughly 9\$ per hour. The protocol was approved by the MIT IRB.

For each tweet, we compare (a) the fraction of right-leaning survey participants out of all survey participants who said that they would consider sharing the tweet with (b) the fraction of users classified as right-leaning that actually retweeted the tweet. We find a positive correlation between the survey responses and the fraction of right-leaning retweeters, Pearson $R=0.5$ ($p = 0.0002$).

\section{Predictive Modeling}
\label{sec:predictive-modeling}
Next, we build models that given the tweet text predict the political diversity of the audience that engaged with the tweet. We discuss our approach for measuring audience diversity, evaluate the accuracy of different machine learning models, and interpret the most accurate model.

\subsection{Measuring Audience Diversity}
\label{subsec:measuring-audience-diversity}
Before discussing the pros and cons of any specific choices, we outline our goals and constraints in measuring audience diversity. First, the main goal is to adopt a measure that will allow us to quantify the extent to which both left- and right-leaning users engaged with a tweet. Second, the measure needs to be intuitive and easy to explain to non-experts, such as the journalists who will use the predictive models to compose new tweets. Third, we need a measure that we can use both in our predictive modeling and when running advertising experiments that test whether choosing tweets with predictive models actually leads to higher audience diversity. 

\textbf{Class Definitions.}
As we will discuss in more detail later, we are much more constrained in what we can measure during the advertising experiments. For instance, we can only specify the set of users who should see the ad and measure their overall engagement. Although we can have more granular measurements of the users’ alignments, we will not be able to know the identities of the individual users who engaged with the content. This constrains us to a definition of audience diversity based on a categorical definition of the users' alignment. As such, we can run separate advertising campaigns for each category of users, measure their engagement, and calculate the diversity. 

The most natural way to categorize users is by whether they are left- or right-leaning. This gives us an intuitive way to measure the diversity of a group of users: the group is most diverse if there is an equal number of left- and right-leaning users and least diverse if the group consists of only left- or only right-leaning users.

Alternatively, we can classify the users into more granular classes, e.g., far left, left, center, right, far right leaning. The benefit of doing so is that we preserve information from the continuous-valued alignment scores.
However, such classification complicates the definition and interpretation of a diversity measure. First, it is unclear how to define diversity under this classification. Should a group of only centrist users have the same diversity as a group with an equal number of left- and right-leaning users? Should we weigh far-left and far-right leaning users more than merely left- and right-leaning users and by how much? Is a group of an equal number of far-left, left, and centrist users more diverse than a group of an equal number of left, centrist, and right-leaning users? Second, even if we answer these questions, explaining and interpreting such a definition is complicated by having to justify subjective choices made is deciding upon the categories.

With these trade-offs in mind, we opt for the binary classification of the users into left- and right-leaning categories. 

\textbf{Diversity Measure.} 
Given the binary classification of the users to left- and right-leaning, one way to measure the diversity of a group of users is to define a discrete random variable $X$ that takes two values, \textit{left} and \textit{right}, with respective probabilities $\pl$ and $\pr$, and compute its entropy. The entropy is commonly used to measure diversity and is often referred to as the Diversity Index or Shannon Index. It is defined as:
\[ H(X) = - \pl \log_2(\pl) - \pr \log_2(\pr). \]
It is maximized ($H=1$) when $\pl = \pr$ and minimized ($H=0$) when $\pl = 0$ or $\pr = 0$.

To estimate $\pl$ and $\pr$ for each tweet, we use Maximum Likelihood Estimation with Laplace Smoothing, i.e., we add a pseudocount of one to the number of observed retweets by left- and right-leaning users:
\[ 
    \pl = \frac{\nl + 1}{\nl + \nr + 2},
    ~
    \pr = \frac{\nr + 1}{\nl + \nr + 2},
\]
where $\nl$ and $\nr$ is the number of observed retweets by left- and right-leaning users, respectively.

This estimation approach also has a Bayesian interpretation: it is equivalent to using a \textit{Beta}(1,1) distribution as the conjugate prior for the parameters of a Binomial distribution. The smoothing has the most significant effect on the estimates of tweets with a small number of retweets, and its effect diminishes as the number of observed retweets increases. Note that we still consider only tweets with at least three retweets. From a practical perspective, the smoothing allows us to distinguish between polarizing tweets with a few retweets (e.g., $\nl=3$, $\nr=0$) and polarizing tweets which have a lot of retweets (e.g., $\nl=100$, $\nr=0$); in the latter case, we have much more information that suggests that the tweet is indeed very polarizing.

\subsection{Learning Methods}
\label{subsec:leaning-methods}
Next, we build models that, given the tweet text, predict the political diversity of its audience. We consider a wide variety of models, from simple linear models on TF-IDF representations of the input text to recent neural network approaches for natural language processing.

Before we apply the models, we preprocess the input text (i.e., the outlets' tweets) by converting them to lower-case, removing punctuation, and replacing numbers with ``\#''. We also remove any URLs and @mentions; however, we keep hashtags as they may carry important semantic information.

\textbf{TF-IDF + Linear Models.}
We start with simple linear models. We consider three ways of representing the text in vector space. (1) We tokenize the (preprocessed) text and build a vocabulary of all uni-grams or all uni-grams and bi-grams. (2) We build a vocabulary of all character n-grams of size three to five, either by including white spaces or respecting the token boundaries. (3) We also test a more sophisticated tokenization technique called SentencePiece~\cite{kudo-2018-subword}, which breaks up tokens into sub-token units selected by analyzing the full corpus. This approach allows us to handle unseen tokens and to control the vocabulary size. Regardless of how we build the vocabulary---uni-grams/bi-grams, character n-grams, or sentence pieces---we always limit the vocabulary size to 32,000. 

Next, given the vocabulary, we encode the tweets using TF-IDF feature representations. We also test whether standardizing the features helps. We scale the features to unit variance but do not center them in order to avoid breaking the sparsity structure of the data.

We test five model types: Linear Regression, Ridge Regression, Lasso, Elastic Net, and Support Vector Regression. For each model, except Linear Regression, we tune the strength of the L1 / L2 regularization parameter, $\lambda \in \{10^{-4}, 10^{-3}, 10^{-2}, 10^{-1}, 10^0, 10^1, 10^2\}$, and train Elastic Nets with equal weight of the L1 and the L2 regularization.

\textbf{Word Embeddings.}
We also consider models based on pre-trained word embeddings. We use the word2vec embeddings, trained on 6B tokens of Google News articles using language modeling as a training objective~\cite{mikolov2013distributed}. 
To obtain tweet embeddings, we consider two ways of aggregating the word embeddings. (1) We average the embeddings of the tweet's words. (2) We use self-attention: we compute an average of the word embeddings weighted by the words' attention scores, which we learn using a two-layer neural network followed by a softmax~\cite{lin2017self}.

After we aggregate the word embeddings, we feed the tweet representation into a series of fully-connected layers with ReLU activations, followed by a prediction layer. We tune several aspects of the learning procedure: whether we freeze or fine-tune the word2vec embeddings, the size of the layers in the attention network (64, 128, 256), and the number (0, 1, 2) and the size (128, 256, 512) of the fully-connected layers. 

\textbf{Recurrent Neural Networks.}
RNNs are particularly suitable for natural language processing as they can be used to encode sequences of arbitrary length and capture dependencies between the tokens in a sequence. We consider two types of RNN architectures: LSTMs and GRUs. In both cases, we train bi-directional models and use the pre-trained word2vec token embeddings as input.

RNNs output a representation/embedding of each token as they process the sequence. We consider three ways of aggregating the embeddings to obtain a tweet embedding. (1) We concatenate the last outputs of the RNN in both directions, left-to-right and right-to-left. Since the RNNs capture contextual information as they process each token, we expect the final outputs to capture longer dependencies. (2) We compute the mean embedding of the RNN outputs for each token. (3) We use self-attention~\cite{lin2017self} and compute the mean embeddings of the RNN outputs weighted by the learned attention scores. Once we aggregate the token embeddings, we feed the tweet embedding into a series of fully-connected layers with ReLU activations. 

We tune the network architecture by testing how the performance changes as we vary the size (128, 256, 512) and the number of RNN layers (1, 2, 3, 4), the pooling mechanism (last embedding, mean embedding, attention with different parameters of the attention network) and the number (0, 1, 2), size (128, 256, 512) and dropout rate (0.0, 0.3) of the fully-connected layers. 

\textbf{BERT.}
Bidirectional Encoder Representations from Transformers or BERT~\cite{devlin2018bert} is a language representation model that processes tokens in relation to all other tokens in the sentence, unlike RNN-based models that process tokens in order, one token at a time. BERT has been used to achieve state-of-the-art results in numerous NLP benchmarks and has been integrated into Google Search, leading to significant improvements in understanding and ranking search queries~\cite{nayak2019understanding}.

To adopt BERT for our task, we average the token embeddings of the last Transformer layer and add a fully-connected layer with a dropout of 0.1 and ReLU activations, followed by a prediction layer. We initialize the network with the pre-trained BERT model and fine-tune it using our dataset. We use the Adam optimizer and batch size of 32 as recommended in the original paper.

\subsection{Experimental Protocol}
To train the models, we split the dataset into 80\% training, 10\% validation, and 10\% test sets using stratified random sampling to preserve the same distribution of tweets per outlet. Due to the long training times of some models, we were unable to run cross-validation. To tune the model architectures and hyper-parameters, we train the models on the training set, evaluate the model variations of the validation set to choose the best model, and measure its performance on the test set. Due to the large number of hyper-parameter combinations and long training times, it was too computationally expensive to perform a grid search. Instead, we first optimized one parameter at a time to find the most promising values and then tested the combinations of those values. 

To train the neural models we evaluate the performance on the validation set after every epoch and stop training and select the last best model if the performance has not improved in the last 5 epochs for BERT and 10 epochs for all other models or if we have reached the maximum number of epochs, 15 for BERT and 100 for all other models. We train the networks using mini-batches of size 32 for BERT and size 64 for all other models. To prevent the gradients from exploding, we clip them to a unit L2 norm after every mini-batch. We use the Adam optimizer setting $\beta_1=0.9$, $\beta_2=0.999$, and L2 weight decay of $0.01$. We consider the following learning rates for BERT, $lr \in \{2 \times 10^{-5}, 3  \times  10^{-5}, 4  \times  10^{-5}, 5 \times 10^{-5}\}$ as recommended in~\cite{devlin2018bert}, and a larger set of values for all other models, $lr \in \{10^{-1}, 10^{-2}, 10^{-3}, 10^{-4}, 10^{-5}, 10^{-6}, 10^{-7}\}$. We use Mean Squared Error as a loss function for all neural models. 
We use Normal approximation to compute confidence intervals of the error rates, but note that bootstrapping the test set and calculating the 2.5 and 97.5 percentile leads to very similar results.

\begin{figure}
\centering
\includegraphics[width=\linewidth]{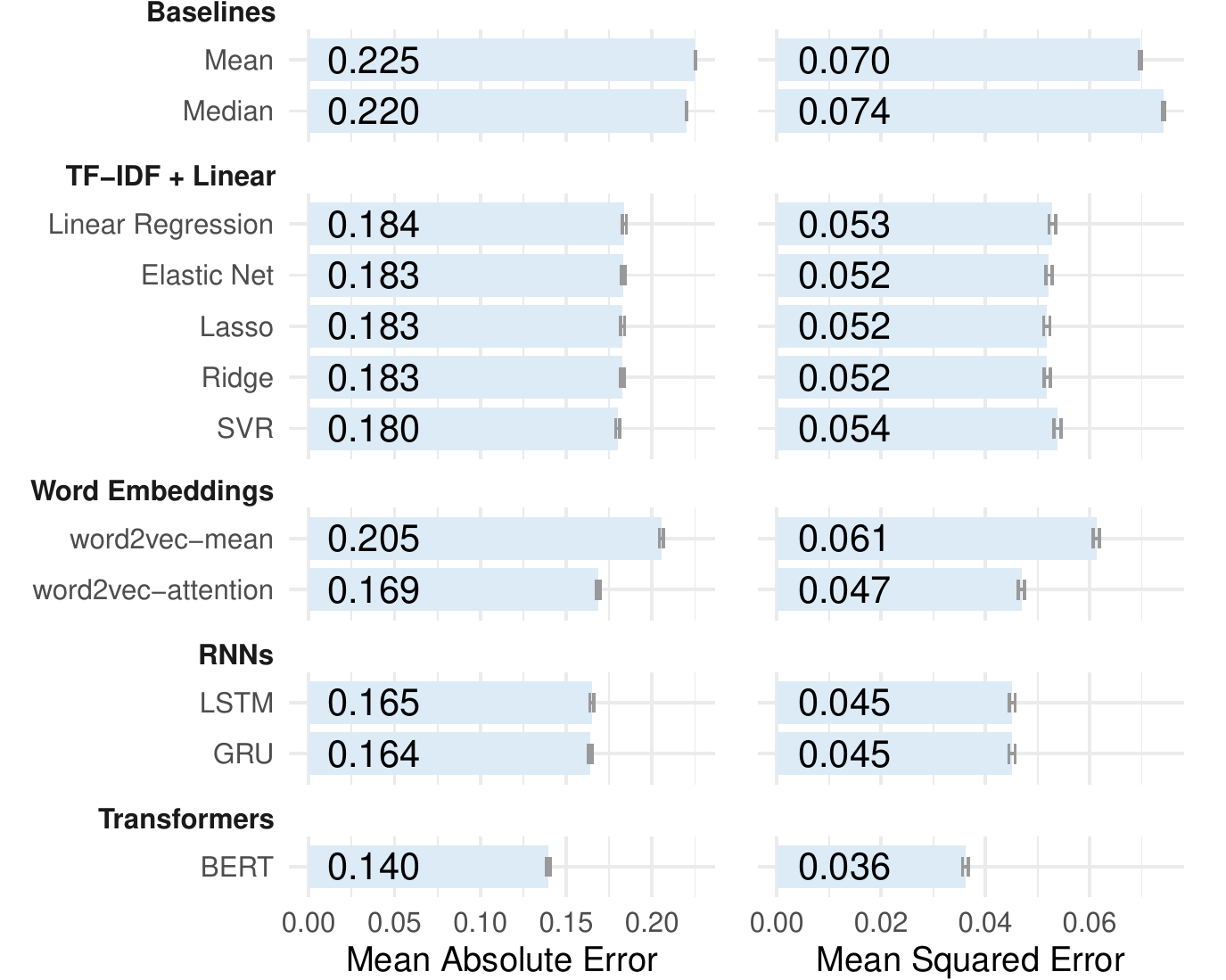}
\caption{Test performance of the regression models predicting the tweets' audience diversity given the tweet text. The error bars represent 95\% confidence intervals.}
\label{fig:modeling-results}
\end{figure}

\subsection{Results}
In Figure~\ref{fig:modeling-results}, we show the Mean Absolute Error (MAE) and the Mean Squared Error (MSE)  of the models on the test set using the hyper-parameters that performed best on the validation set. To put the results in perspective, we use two constant predictors as baselines: the mean ($\mu=0.44$) and the median ($\mu_{\frac{1}{2}}=0.37$) of the tweets’ audience diversity in the training set. The mean minimizes the MSE, and the median minimizes the MAE of the target variable in the training set.

We find that all linear models trained on the TF-IDF representations of the tweet text perform significantly better than the baselines. All models have a similar performance with MAE ranging between 0.18 and 0.184. We observe that tokenizing the text using SentencePiece tokenization and including all uni-grams and bi-grams in the vocabulary leads to the best performance for all linear models except for SVR, which works slightly better with regular tokenization. 

We find that the mean word embedding of the tweet tokens is not a good predictor of the tweet audience diversity. In fact, the models based on averaging the word2vec embeddings perform worse than the linear models trained on TF-IDF features. However, using self-attention, i.e., learning different weights for each word embedding, leads to significantly better results, improving over the linear models. 

We observe that the RNNs work slightly better than the word2vec embeddings with self-attention. The two variants, GRUs and LSTMs, achieve very similar results. Both models perform best when aggregating the RNNs outputs using self-attention and using a single RNN layer.

Finally, we find that the fine-tuned BERT model performs best, significantly outperforming the RNN models. It achieves a MAE of 0.14 and a MSE of 0.036. We observed that the model performs well with different learning rates and different sizes of the final fully-connected layer that we added to the network architecture. We note that the BERT model also performs best when we evaluate the predictive performance for each outlet individually. We use this model for all subsequent analyses.

\subsection{Model Interpretation}
\label{subsec:model-interpretation}
Next, we examine the fine-tuned BERT model that achieves the best out-of-sample prediction performance. We use a popular model interpretation method to attribute predictions to the presence of specific input words.  Aggregating these {\it attribution values} across inputs gives us insight about what kinds of words are most aligned with audience diversity.  We examine several lexical categories based on part of speech, subjectivity, sentiment level, and emotional type.

\textbf{Word Attribution Values Using Integrated Gradients.}
While highly predictive, deep neural models like BERT are known to be difficult to interpret. To attribute the model predictions to input words, we used Integrated Gradients~\cite{sundararajan2017axiomatic}. This method has two attractive properties: (1) sensitivity, i.e., any change of an input feature in a single instance that changes the model prediction is given a non-zero attribution value; and (2) implementation invariance, i.e., the attribution values of models that produce the same outputs for all inputs are always the same, regardless of their implementation. Integrated Gradients computes an attribution value for each feature in an input instance that represents how much its presence changed the model prediction relative to other features.\footnote{As recommended in~\cite{sundararajan2017axiomatic}, we use a sequence of padding tokens as a reference input to represent the absence of a signal.} A positive attribution indicates that the input feature is positively correlated with the model output for the given instance and a negative attribution indicates a negative correlation. 

As BERT splits the input text into word-pieces instead of words, we sum the attribution values of the constituent word-pieces to obtain attribution values for whole words. To quantify the overall influence of a word on the model predictions, we compute the mean and confidence interval of its attribution value across all documents in the test set ($N_{test} = $~56.6K). Similarly, to compute the overall influence of a set of words (e.g., positive words), we compute the mean of the attribution values of all words in the set across all documents in the test set.

\begin{figure}
\centering
\includegraphics[width=\linewidth]{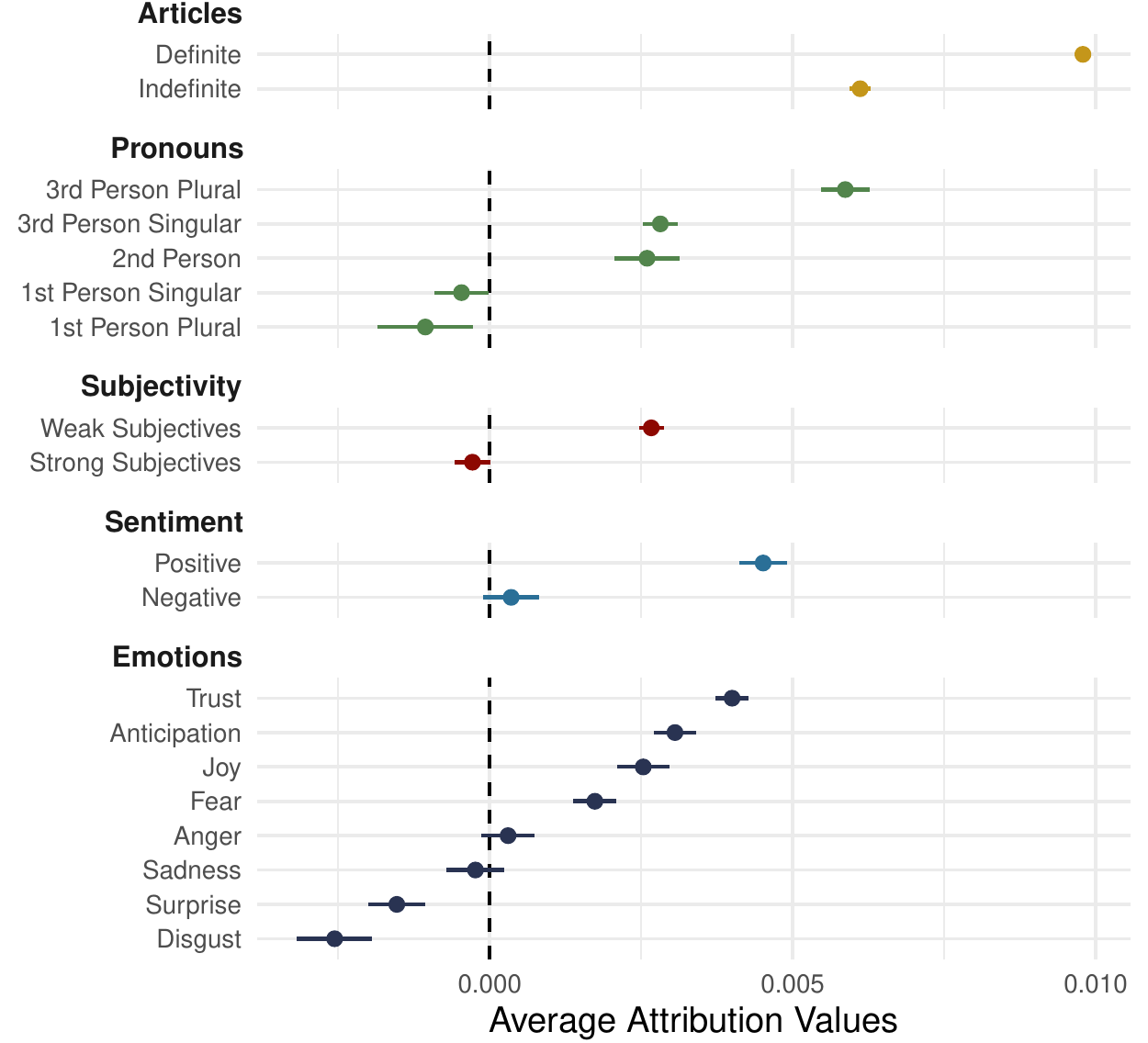}
\caption{Analysis of the attributions of the model predictions to the presence of input words from five lexical categories relative to the absence of a signal. The error bars represent 95\% confidence intervals over the test set.}
\label{fig:attribution-results}
\end{figure}

\textbf{Attribution Analyses.}
Next, we investigate the attribution values of various word categories (Figure~\ref{fig:attribution-results}).

\mybullet 
\textit{Articles.} We start by comparing the attribution values of definite (the) vs.\ indefinite articles (a, an). Previous studies have found that indefinite articles are more frequent than definite articles in viral tweets~\cite{tan2014effect} and memorable movie quotes~\cite{danescu2012you}. In contrast, we find that, on average, definite articles have larger attribution values than indefinite articles, suggesting that specificity in the posts’ language is correlated with higher audience diversity. 

\mybullet 
\textit{Pronouns.} To examine the attribution values of personal pronouns, we use the pronouns listed in the LIWC lexicon~\cite{pennebaker2015}. Past work has found that first-person pronouns occur more often in successful novels~\cite{ashok2013success} and that third-person pronouns occur more often in viral tweets~\cite{tan2014effect}. We find that the attribution values of first-person pronouns are negatively correlated, while the attribution values of second- and third-person pronouns are positively correlated with predictions of high audience diversity. 

\mybullet 
\textit{Subjectivity.} We investigate the impact of strongly vs.\ weakly subjective words using the MPQA subjectivity lexicon~\cite{wilson2005recognizing}. Strongly subjective words are those that are subjective in most contexts, while weakly subjective words are those that are subjective only in some contexts. Recasens et al.~\citeyearpar{recasens2013linguistic} find that the presence of both weakly and strongly subjective words is predictive of bias in Wikipedia articles. We find that weakly subjective words are positively, while strongly subjective words are slightly negatively correlated with predictions of high audience diversity. 

\mybullet 
\textit{Sentiment.} We analyze the influence of words with positive vs.\ negative sentiment on the model predictions. We use the lists of positive and negative emotion words in LIWC~\cite{pennebaker2015}, but note that using the MPQA~\cite{wilson2005recognizing} and NRC~\cite{mohammad2013crowdsourcing} sentiment lexicons leads to the same conclusions. Previous studies have found that positive news articles are more popular than negative ones~\cite{berger2012makes} and that viral tweets are more likely to contain both positive and negative words~\cite{tan2014effect}. We find that positive words have positive attribution values, while negative words have attribution values close to zero, suggesting that the presence of positive words leads to predictions of higher audience diversity. 

\mybullet 
\textit{Emotions.} Finally, we use the NRC~\cite{mohammad2013crowdsourcing} emotions lexicon to analyze the attribution values of words associated with specific emotions. We find that words that evoke trust, anticipation, joy, and fear have positive attribution values, words that evoke anger and sadness have attribution values close to zero, and words that evoke surprise and disgust have negative attribution values.

In summary, we find that posts that contain definite articles, second- and third-person pronouns, less subjective, and more positive words are seen as more likely to have higher audience diversity by the model.

\section{Web Application}
\label{sec:news-bridge-web-app}
To make the models more easily accessible to Frontline’s journalists, we build a web application that surfaces the model predictions. The goal of the application is to allow the journalists to quickly iterate on tweet drafts based on the model predictions and to help them select candidate tweets from film transcripts. 

\subsection{Main User Interface}
Figure~\ref{fig:news-bridge-main} shows a screenshot of the main page of the application. The users can enter draft tweets in the input box (Figure~\ref{fig:news-bridge-main}\#1), press submit, and get the model predictions in the results table. The input can consist of multiple tweets, separated by a new line, and each tweet can consist of multiple sentences that will be scored together. 

The results table (Figure~\ref{fig:news-bridge-main}\#2) shows the input text and the predicted ``bridginess'' score, a user-friendly name for the audience diversity measure based on the entropy of the distribution of left- vs.\ right-leaning users (Section~\ref{sec:predictive-modeling}). The color of the table cells containing the scores varies from light-green for non-bridging tweets to dark-green for bridging tweets. 

Based on the journalists' feedback, we also added tweet alignment score predictions. The goal of these scores is to supplement the bridginess scores. To make the alignment predictions, we trained a BERT model equivalent to the one we used to make the bridginess predictions, but instead of the entropy, we used the retweeters' average political alignment score as a target.  Similar to the bridginess scores, we vary the color of the table cells that contain the alignment scores on a gradient from blue (if the score is negative, i.e., the tweet is predicted to be more engaging to left-leaning users) to red (if the score is positive).

\subsection{Explanations}
After the initial user tests of the tool, the main feedback we received was that while the scores are informative, it is often unclear why the model makes the predictions it does. To address this, we show two kinds of explanations to supplement the predictions: (1) we highlight certain words in the tweets and display relevant corpus statistics, and (2) we show historical tweets that are semantically similar to the input tweet and have a high bridginess score.

\begin{figure}[t]
\centering
\includegraphics[width=0.98\linewidth]{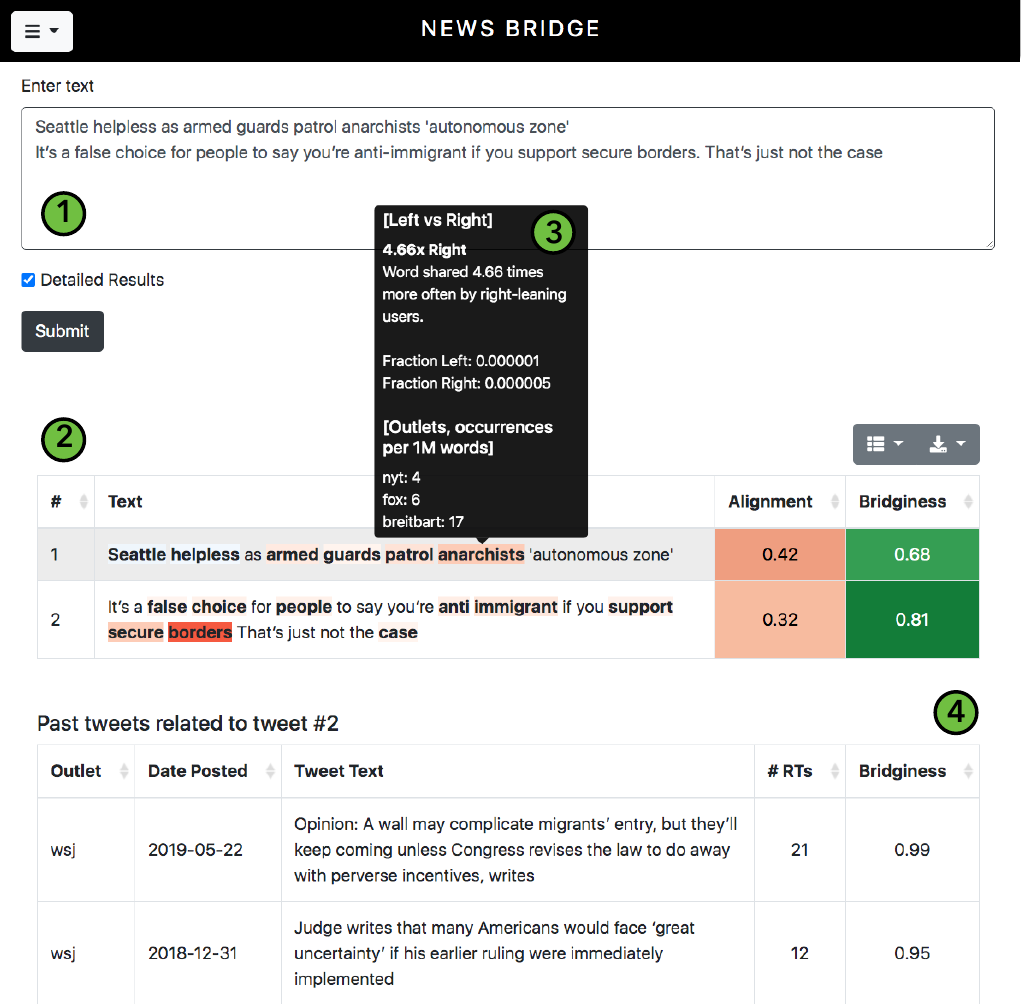}
\caption{Screenshot of the web application's user interface.}
\label{fig:news-bridge-main}
\end{figure}

\textbf{Word Highlighting.}
We compute how often each word is retweeted by left- and right-leaning users. 
More specifically, we compute the probability $\pwl$ (and $\pwr$) as the ratio between the number of times the word appears in retweets by left- (right-) leaning users and the total number of words in all retweets by left- (right-) leaning users. 
We highlight the words of the input text in blue or red depending on whether they are more likely to be retweeted by left- or right-leaning users, and we set the brightness of the color in proportion to the ratio between the two quantities ($\pwl$ and $\pwr$). We also compute how often each word appears in tweets posted by each of the five media outlets and Frontline. When the user hovers over the word, a pop-up shows the different word statistics (Figure~\ref{fig:news-bridge-main}\#3). 

Beyond highlighting words based on simple word statistics, we also considered two other, more sophisticated approaches. (1) A common way of visualizing which words in the text were most important in the model prediction is to use the self-attention weights. However, recent studies have shown that attention weights do not provide meaningful explanations for the model predictions~\cite{jain2019attention}. 
(2) We also considered using Integrated Gradients---the method we used in Section~\ref{subsec:model-interpretation}---but found that it takes about 10 to 20 seconds to compute the attributions on a single prediction. Since the goal of this tool is to allow journalists to quickly iterate on the tweet text based on the model predictions, we decided that increasing the latency of the predictions would significantly degrade the user experience.

\textbf{Similar Historical Tweets.}
One of BERT's main advantages is that it models the relationships between all the words in the sentence together. As a result, highlighting individual words is unlikely to fully explain its predictions. Therefore, in addition to providing word statistics, we also show similar historical tweets that were bridging. The goal is to show the user sample tweets that look similar to the model but have a higher bridginess score. To represent the tweets, we use the embeddings generated in the last layer of BERT. We save the embeddings of all tweets in the dataset, and given the embedding of the input tweet, we find the nearest neighbors in the embedding space. To index and search the tweet embeddings efficiently, we use \textit{Faiss}~\cite{johnson2019billion}, a library for similarity search of dense vectors. When the user clicks one of the rows in the results table, we show the ten most similar tweets to the input tweet, including when they were posted, by which outlet, how many retweets they received, and their bridginess score (Figure~\ref{fig:news-bridge-main}\#4).

\subsection{Transcript Analysis}
To streamline the selection of bridging tweets, we also analyze the transcripts of the Frontline documentaries and show the results through an interactive interface. Frontline’s tweets often include quotes from the documentaries, and our models can help with the selection of quotes or film segments that might be engaging to a politically diverse audience. Since the models are trained on a different domain, tweets vs.\ transcripts, the goal is not to select transcript segments to be posted as promotional tweets but rather to help identify segments that can serve as a good starting point for composing bridging tweets.

To analyze each transcript, we parse the transcript segments and use the BERT models to predict the expected bridginess and alignment scores of each segment. To provide an overview of the predictions, we plot the scores against the segment number and show a table of the results that includes the segment number, the speaker, the segment text, and the predicted scores. 
We received very positive feedback from the journalists about this feature of the web application.

\section{Advertising Experiments}
\label{sec:ads-exps}
Next, we test whether the predictive models we developed can be effectively used to compose tweets that engage a more politically diverse audience. In partnership with Frontline, we ran ten advertising experiments on Twitter between May and August of 2020. In each experiment, we selected a pair of tweets---one predicted to be engaging to a more (treatment) and one predicted to be engaging to a less (control) diverse audience by our model---and measured the engagement of left- and right-leaning users with each tweet. While the advertising experiments presented in this section are not randomized experiments, they are the only way to run experiments on the platform and measure how thousands of Twitter users respond to the test tweets. In this section, we discuss how the advertising campaigns were set up, explain how the target audiences and the test tweets were selected, and summarize the results of the experiments.

\begin{figure}
\includegraphics[width=\linewidth]{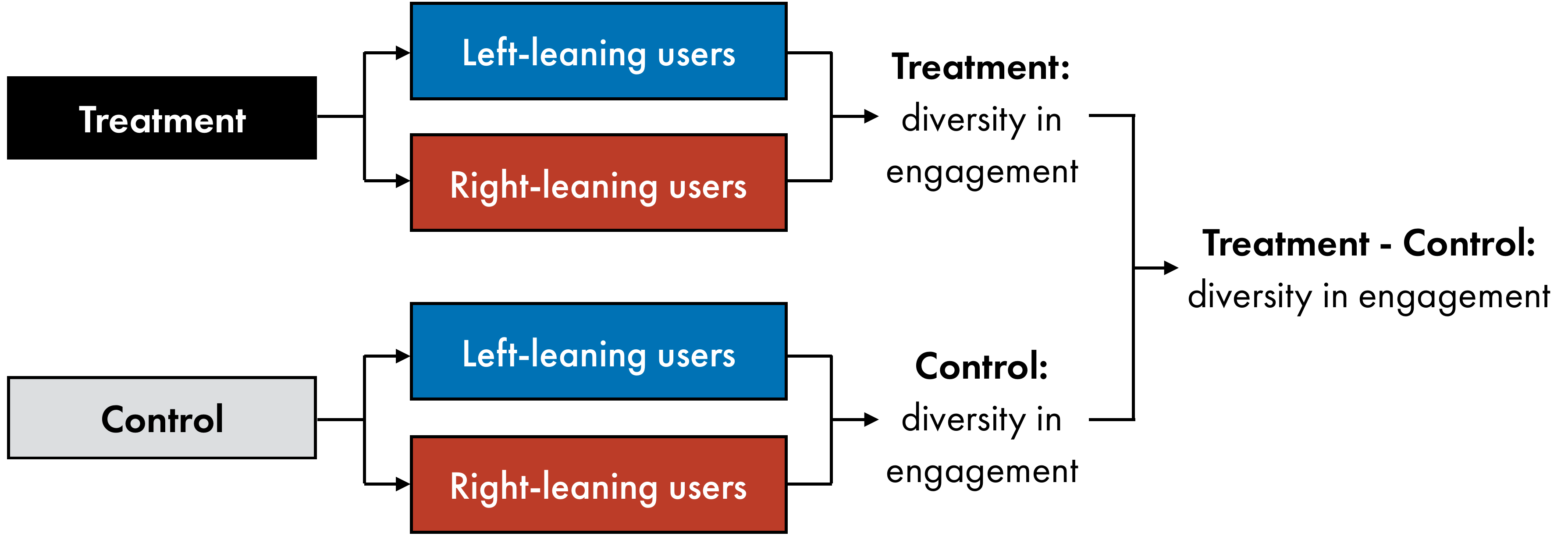}
\caption{Advertising experiments setup. The treatment and control are tweets predicted to be engaging to more and less politically diverse audiences, respectively. The tweets are selected using the predictions of the best-performing~model.}
\label{fig:ad-exps-sketch}
\end{figure}

\subsection{Campaign Setup}
We start by describing the setup of the advertising experiments. One of the challenges in measuring the audience diversity of the test tweets is that the Twitter advertising platform only reports aggregate metrics of engagement and does not report the engagement levels by users with different political leanings. To address this issue, we run two campaigns for each test tweet---one targeting only left-leaning users and another targeting only right-leaning users---and measure the aggregate engagement of each group. Thus, in each experiment, testing a pair of tweets, we run four campaigns in total (Figure~\ref{fig:ad-exps-sketch}).

We run all experiments for five days, Wednesday to Sunday, since Frontline airs new documentaries on Tuesday evenings.\footnote{Except for one experiment, which we had to relaunch two days later after discovering a misspelling in one of the test tweets.}
To avoid any time-related confounders, we schedule all four campaigns in each experiment to start and end at the exact same time. The Twitter advertising platform requires us to specify a campaign objective, which is aimed to help advertisers maximize their ROI. We use ``awareness'' as a campaign objective which, unlike other objectives, optimizes for reach and does not explicitly optimize engagement. We set this objective to minimize the interference of the advertising engine and ensure that the test tweets are not shown only to users who are likely to engage with them.

We set the budget for each of the four campaigns to \$250 and capped the daily budget to \$50 to spread out the experiments. To make sure that the test tweets are shown to as many of the targeted users as possible, we bid at \$15 / 1,000 impressions, significantly higher than the range recommended by the advertising platform, \$3.5--\$6 / 1,000 impressions. Also, to ensure that users in one group are not exposed to the advertisements more often, we limit the number of impressions per user to one.

To ensure that users are not exposed to both the treatment and control tweets, we used so-called ``promoted-only tweets''. These tweets are shown only to the users selected to see the advertisements and do not appear on the account page (i.e., @frontlinepbs), the followers' timelines, or the search results. Promoted tweets are the same as organic tweets in every aspect except that they have a ``Promoted’’ label at the bottom of the tweet. 

To ensure that the test tweets are shown only to left-leaning or only to right-leaning users, we target so-called ``tailored audiences''. Rather than specifying the target audience using audience characteristics (e.g., age: 25-49), the tailored audiences allow us to specify a custom set of users by uploading a list of user ids. Next, we discuss how we selected the custom audiences.

\subsection{Audience Selection}
\label{subsec:ad-exps-audience-selection}
To select the users who will be exposed to the test tweets, we consider only the followers of @frontlinepbs and the followers’ followers (i.e., two hops away from the Frontline account\footnote{Due to the limits of the Twitter API, we do not include the followers of @frontlinepbs followers with more than 5,000 followers. They constituted only 3.85\% of all @frontlinepbs followers.}). 
Initially, we included only Frontline followers, but to ensure that enough users are eventually exposed to the tweets, we had to expand this set. Out of these users, we further restrict to users whose political alignment we can estimate based on their content-sharing patterns. 

We randomly select 200K users for each experiment, 100K left- and 100K right-leaning users, and, within each group, we randomly assign half of them to treatment and half of them to control (treatment-left: 50K, control-left: 50K, treatment-right: 50K, control-right: 50K). We ensure that an equal proportion of @frontlinepbs followers vs.\ followers of followers is assigned to each treatment arm. 

Once we make the assignments, we perform a number of balance checks to ensure that there are no systematic differences between users in different groups. We test for balance between three pairs of groups: (1) left-leaning users in treatment vs.\ control, (2) right-leaning users in treatment vs.\ control, (3) all users in treatment vs.\ all users in control. We consider the following user characteristics: number of posts, likes, followers, friends, tenure on Twitter, and the numerical estimate of their political alignment. We log the number of posts, likes, friends, and followers, since their distributions are highly skewed. We run two types of covariate balance analysis. (1) We regress the user characteristics on the treatment assignment using logistic regression and ensure that none of the coefficients are statistically significant. (2) We use a permutation test, i.e., we compare the log-likelihood of the logistic model regressing the user characteristics on the treatment assignment with its empirical distribution under random reassignments of treatment that follow the same randomization scheme~\cite{gerber2012field}. To obtain the empirical null distribution, we measure the log-likelihood of 10K reassignments.

Once we have selected the audiences, we upload them to the Twitter advertising platform. While we upload a list of 50K user ids, only 15K--18K (i.e., 32\%--37\%) of them could be targeted. According to the documentation, inactive users are excluded from the tailored audience, but it is unclear what criteria are used to determine whether a user is active. Furthermore, when we use the tailored audiences to run a campaign, only 8.5K--9K (i.e., 17\%--18\%) of the users are shown the test tweets. There might be several factors that lead to this: (1) the users may simply be less active and  not have used Twitter when we ran the campaign, (2) the users might be very desirable and many other campaigns may have bid for them, or (3) Twitter’s algorithm has predicted that they are less likely to engage with the tweet and has given less priority to our campaign. There could be other factors we are not aware of. Since we have no control over who will be exposed to the test tweets, this is where our experiments depart from traditional A/B tests.

We note that left-leaning users are both more likely to be part of the tailored audience and more likely to be shown the test tweets. Among other reasons, this might be because they are more active or because the advertising engine predicts that they have a higher affinity to engage with Frontline's content. While there are differences between the left- and right-leaning subgroups, the treatment and control groups as a whole are very similar (more details in Section~\ref{subsec:ad-exps-results}).

\subsection{Content Selection}
All test tweets were composed by Frontline's journalists using the web application (Section~\ref{sec:news-bridge-web-app}) which surfaces the predictions of the best-performing model (Section~\ref{sec:predictive-modeling}). To select the tweets, they relied on the predicted audience diversity score, which we refer to as the ``bridginess'' score in the web application (Figure~\ref{fig:news-bridge-main}). While we administered the advertising campaigns, we were not involved in the selection process and only provided guidance on how to use our tools. 

We instructed the journalists to ensure that the difference in the audience diversity predictions between treatment and control tweets in each experiment is at least 0.1. Across the ten experiments, the average difference of the model predictions between the treatment and control tweets was 0.25 (min: 0.1, max: 0.4). In any single experiment, the treatment and control tweets were composed by the same journalist.

To avoid any topical confounders, the pair of treatment and control tweets in each experiment was about the same documentary. All test tweets included a link to the relevant documentary and a promotional image. We used the same promotional image in both the treatment and control tweets in each experiment. To maximize the integrity of the experiments, we ensure that none of the test tweets were previously published, i.e., posted on Twitter or any other social media platforms. 

The tweets posted by Frontline go through the same editorial scrutiny as other content published by Frontline and need to be approved before publication. All tweets were approved by the same editorial board. The timing of the experiments was also determined by Frontline's schedule. As we mentioned earlier, each pair of tweets was tested during the same time frame.

\subsection{Results}
\label{subsec:ad-exps-results}
Next, we analyze the results of the experiments. We start by establishing the validity of the experimental design. We compare the aggregate statistics of the users that were exposed to the treatment and control tweets in each of the ten experiments. We consider three key user characteristics reported by the Twitter advertising platform: age (18-24/25-49/50+), gender (male/female), and language (English/other). We do not find any systematic differences between (1) the users exposed to the treatment and control tweets ($p=0.9$, permutation test, descibed in Section~\ref{subsec:ad-exps-audience-selection}), (2) the left-leaning users exposed to the treatment and control tweets ($p=0.4$), or (3) right-leaning users exposed to the treatment and control tweets ($p=0.3$). These results suggest that differences in the observed diversity of engagement between the treatment and control tweets are not due to an imbalance between the two groups.

Next, we turn to the main outcome of the experiments. We measure the diversity of the audience engagement of the treatment and control tweets using the definition explained in Section~\ref{subsec:measuring-audience-diversity}. We focus on overall engagement and do not analyze the number of likes and retweets individually as there are too few such interactions to make meaningful comparisons. 

\begin{figure}
\centering
\includegraphics[width=\linewidth]{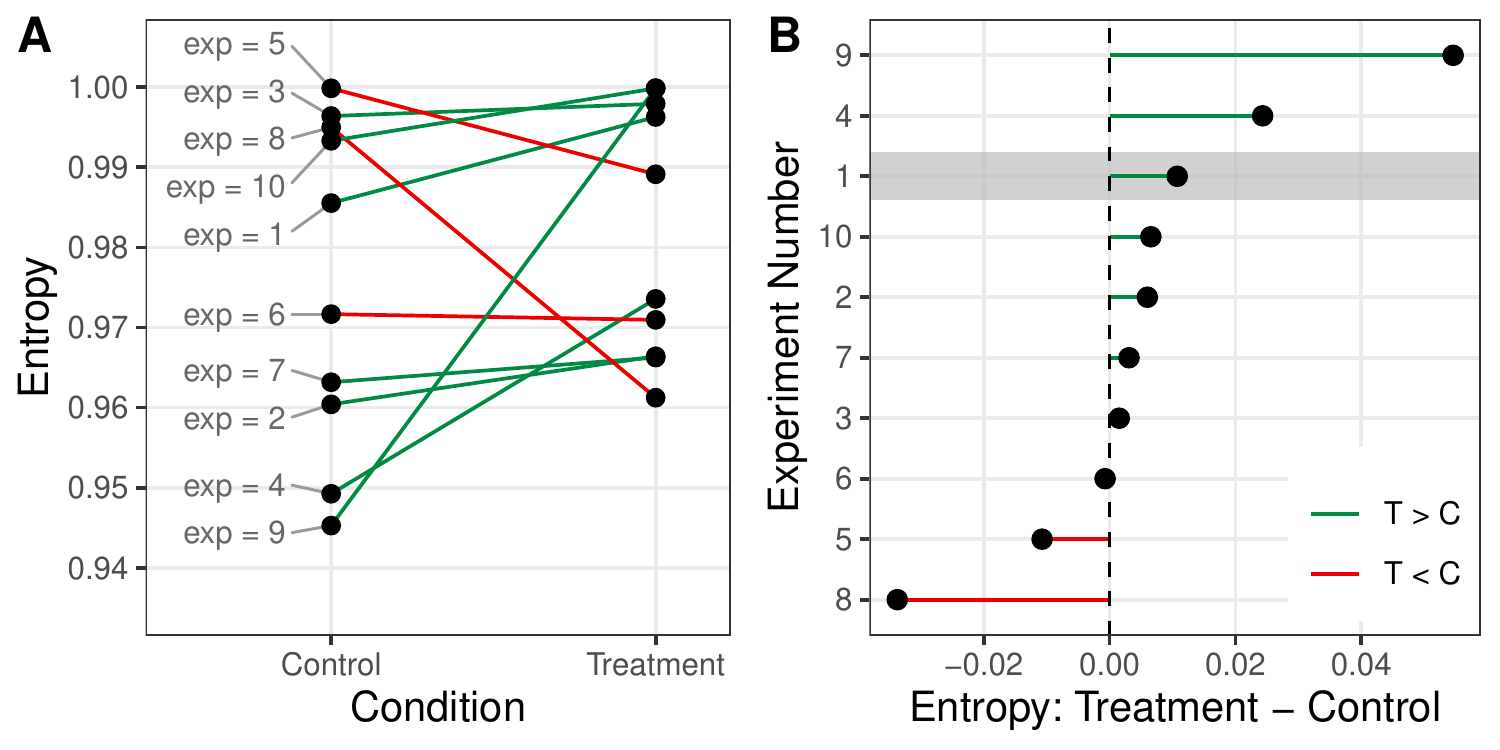}
\caption{Comparison between the audience diversity of the treatment and control tweets in the advertising experiments.}
\label{fig:ad-exps-main}
\end{figure}

We find that in seven out of the ten experiments, the treatment tweets achieved higher audience diversity than the control tweets, matching our model's predictions (Figure~\ref{fig:ad-exps-main}). 
The average difference in audience diversity between the treatment and control tweets is 0.0062 ($p=0.22$, computed using randomization inference~\cite{gerber2012field}).

As differences in entropy might be hard to interpret, it is useful to examine the differences in engagement between left- and right-leaning audiences for a pair of treatment and control tweets to contextualize the results. For instance, in \textit{Experiment 1} (highlighted in Figure~\ref{fig:ad-exps-main}B), the difference in entropy of 0.0108 corresponds to a two-fold reduction in the difference between the probability of engagement by left- and right-leaning audiences in the treatment vs.\ control tweets, i.e., $\DeltaT = 0.072$ ($\plT = 0.536$, $\prT = 0.464$), vs.\ $\DeltaC = 0.142$ ($\plC = 0.571$, $\prC = 0.429$).

We observe that across the ten experiments, both the treatment and the control tweets have a high audience diversity, i.e., entropy values close to one (Figure~\ref{fig:ad-exps-main}A). 
This is partly due to the shape of the entropy function:  around 0.5, small changes in the balance between left- and right-leaning users ($\pl$ and $\pr$) lead to even smaller increases in entropy. 
For instance, if the breakdown of engagement with the control tweet is $\pl=0.45$ and $\pr=0.55$ (entropy~$=$~0.9927) and the treatment tweet achieves perfect balance in audience engagement, i.e., $\pl=0.5$, $\pr=0.5$ (entropy~$=$~1), then that would be an increase in entropy of only 0.007.

Finally, we investigate the trade-off between audience diversity and overall engagement. We find that, while the treatment tweets engage a more politically diverse audience, they received 10\% fewer clicks than the control tweets ($p=0.14$, standardized permutation test). These results suggest that engaging a more politically diverse audience may come at the cost of lower overall engagement.

In summary, the advertising experiments illustrate the potential of using our approach to engage more politically diverse audiences. However, we underline that additional experiments are needed to make more definitive conclusions about the generalizability of our results.

\section{Further Related Work}
In addition to the previously mentioned literature, we draw on a rich body of research studying media slant, political polarization online, linguistic variation across communities, and causal effects of different lexical choices.

Many studies have sought to quantify media outlets’ bias, slant, or alignment based on the similarity between the language used by the outlets and that used in congressional proceedings~\cite{gentzkow2010drives}, crowdsourced annotations~\cite{budak2016fair}, audience demographics extracted from advertising platforms~\cite{ribeiro2018media}, and social media sharing patterns of self-identified partisans~\cite{bakshy2015exposure}.

Several recent studies have examined how users' political leaning influences their perceptions of advertisements~\cite{ribeiro2019microtargeting} and news stories~\cite{babaei2019analyzing} on social media, finding that left- and right-leaning users markedly differ in their likelihood to agree with, report as inappropriate, or perceive as true the very same content. Another line of work has proposed methods for identifying ``purple'' news tweets---which prompt similar reactions from left- and right-leaning users---based on aggregate statistics of the political leanings of the outlet’s followers and the tweet's retweeters and repliers~\cite{babaei2018purple}.

Researchers have also proposed methods for reaching a diverse audience by strategically seeding information in a social network~\cite{anwar2021balanced, garimella2017balancing} rather than considering the content of the posts. Other studies have tested interventions that nudge users’ behaviors to reduce political polarization by incentivizing users to set their browser homepage to a left- or right-leaning news source~\cite{guess2021consequences}, allowing users to replace their feeds with those of users with opposing views~\cite{saveski2022perspective}, or encouraging users to examine the political homogeneity of their social networks~\cite{gillani2018me}. 

Prior work has investigated the linguistic variation across different online communities, including linguistic differences between Republicans and Democrats~\cite{demszky2019analyzing}, politically homogeneous vs.\ heterogeneous groups~\cite{an2019political}, and overall stylistic variation among different communities on different social media platforms~\cite{khalid2020style}. Several studies have used observational causal inference to estimate the causal effects of different lexical choices. Park et al.~\shortcite{park2021understanding} analyze the differences between news headlines and the corresponding promotional social media posts, and estimate the effects of different editing strategies on audience engagement. Similarly, Wang and Culotta~\shortcite{wang2019words} investigate how single-word substitutions affect the audience’s perception of a sentence. 

Finally, several recent papers have discussed the challenges (Eckles et al.~\citeyear{eckles2018field}) and opportunities~\cite{guess2021experiments} of using advertisements to run social media experiments.

In contrast to these studies, we investigate the relationship between the content of the social media posts and the political diversity of their audience, develop tools that help media outlets reach a more politically diverse audience, and test them using advertising experiments.

\section{Discussion and Conclusion}
\label{sec:conclusion}
In this paper, we investigated the relationship between the content of tweets posted by media outlets and the political diversity of the users who engaged with them. We collected 566K tweets by five media outlets and the documentary series Frontline over more than three years. To measure each tweet's audience diversity, we computed the entropy of the distribution of left- vs.\ right-leaning users who shared the tweet. Using this data, we trained various models that, given the tweet text, predict the audience diversity. We then integrated the best model into a web application designed to help Frontline's journalists craft more bridging tweets, guided by the model predictions. Finally, in partnership with Frontline, we ran ten advertising experiments to test whether the model predictions can be effectively used to compose more bridging tweets. We found that in seven out of the ten experiments, the tweets selected by our model were indeed engaging to a more politically diverse audience, illustrating the effectiveness of our approach.

Studies of political polarization on social media typically investigate the behaviors of individual users. In this work, we focused on the media outlets' role in fragmenting the audience and developed tools that can help journalists reduce it. Our predictive models can be integrated into existing assistive writing technologies that journalists already use to enhance their writing process. From a methodological perspective, our design of the advertising experiments can help other researchers test new interventions on social platforms and achieve greater external validity of their findings.

Nonetheless, our work has limitations that open avenues for future work.
First, since the advertising experiments are expensive and time-consuming---with most of the burden falling on the journalists who select, write, and approve the content---we were unable to run a large number of experiments. While the current results illustrate the effectiveness of our tools, the limited number of experiments prevents us from making definitive conclusions. Running additional experiments to test how our tools perform across different topics over a longer time frame remains a direction for future work. 
Second, while we used different features of Twitter’s advertising platform to design experiments that resemble randomized experiments (used tailored-audiences, promoted-only tweets, capped the number of exposures per user, and placed high bids), we were unable to remove the influence of the advertising engine entirely. More specifically, due to algorithmic predictions or market forces, the advertising engine may show the test tweets to users who are more likely to engage with them, instead of a random subset of users. While this is equally likely to occur in the treatment and control conditions, we cannot rule out the possibility that the higher audience diversity of the treatment tweets is not due to differences in the tweet content but due to differences in the delivery of the advertisements. Running further experiments with randomized assignment administered by the advertising engine when such a feature is available on Twitter is a promising avenue for future work.

This work is an initial step in investigating the relationship between content and the audience’s political diversity on social media and developing tools that help journalists reach a more diverse audience. We hope that our analysis will encourage further work on the relationship between the content and the composition of the audience that engages with it. We also hope that our methodology for running advertising experiments can serve as a guide for academic researchers who want to test various interventions on social media platforms but do not have direct access to their users.

\section*{Acknowledgments}
We are grateful for comments from Lada Adamic, Dean Eckles, members of the Social Analytics Lab at MIT Sloan, Ugander Lab at Stanford, and the anonymous reviewers. We thanks the members of the Frontline team: Katherine Griwert, Ben Abrams, Pam Johnston, and Raney Aronson-Rath. We also thank Twitter for the financial and data support.

\fontsize{9.0pt}{10.0pt}  %
\selectfont
\bibliography{refs}

\end{document}